# Preliminary Studies on the Usefulness of Nonlinear Boundary Element Method for Real-Time Simulation of Biological Organs

Kirana Kumara P
Assistant Professor – Senior Scale,
Department of Mechanical & Manufacturing Engineering,
Manipal Institute of Technology,
Manipal Academy of Higher Education, Manipal, Karnataka,
India – 576104
Email: kiranakumarap@gmail.com

**Abstract:** There is some literature on the application of linear boundary element method (BEM) for real-time simulation of biological organs. However, literature is scant when it comes to the application of nonlinear BEM, although there is a possibility that the use of nonlinear BEM would result in better simulations. Hence the present paper explores the possibility of using nonlinear BEM for real-time simulation of biological organs. This paper begins with a general discussion about using the nonlinear BEM for real-time simulation of biological organs. Literature on nonlinear BEM is reviewed and the literature that deal with nonlinear formulations and coding are noted down next. In the later sections, some results obtained from nonlinear analyses are compared with the corresponding results from linear analyses. The last section concludes with remarks that indicate that it might be possible to obtain better simulations in the future by using nonlinear BEM.

## 1. Introduction

Biological organs are inherently nonlinear. Hence the simulation of biological organs that consider nonlinearity would result in more accurate results. However, whether the simulation would be more realistic after incorporating nonlinearity depends not only on the accuracy of the results obtained but also on how fast the simulations are (whether or not the simulations are real-time, since not just accuracy but the real-time performance also increases realism).

Hence one has to incorporate nonlinearity into the BEM based simulation of biological organs if one wants to answer the questions: "Is it possible to achieve more realistic simulations by incorporating nonlinearity into the BEM based simulation of biological organs? Is it possible to achieve the real-time performance with such models? Is the accuracy offered by such models enough in case the models can only be made of very few elements to achieve the real-time performance? Or, whether one can achieve better realism by using a greater number of nonlinear elements even if it could result in simulations that are not strictly real-time (nearly real-time simulations)?"

Clearly, from the results presented in the previous chapters, one can conclude that it is difficult to achieve the real-time performance if nonlinearity is to be incorporated into the simulations. However, it should be possible to achieve the real-time performance if very few nonlinear boundary elements are used. On the contrary, it may be possible to achieve nearly real-time performance (although not strictly real-time performance), even if the total number of boundary elements is kept the same as those in the simulations carried out in the previous chapters. Whether or not such nearly real-time nonlinear simulations are preferred over the simulations which are strictly real-time but do not incorporate any nonlinearity is a question which can only be answered by competent and experienced surgeons, after they use surgery simulators built by using both these technologies (one technology at a time). But the first step towards building a BEM based simulator that can simulate nonlinearity could be to carry out the nonlinear simulation of biological organs using BEM.

It might be important to note that no literature is available on the simulation of biological organs using nonlinear BEM, whether in the context of real-time simulations or otherwise.

Even when solving nonlinear problems, the BEM usually uses the same fundamental solutions as those used for linear simulations. This might lead to lesser accuracy. Moreover, the BEM formulations that are generally employed to solve nonlinear problems usually require meshing of the interior of the problem domain, not just the boundary of the problem domain. This can make the BEM less attractive (over techniques like FEM) because one of the reasons for choosing the BEM over techniques like FEM is that it requires meshing of only the boundary of the problem domain (at least for linear problems). It may be noted that once the BEM loses this advantage (when solving nonlinear problems), there may not be any advantage in using the BEM over FEM (in terms of speed and accuracy). Of course, still there is a need to carry out the simulation of biological organs by using the BEM and FEM both and find out which of the numerical techniques is better suited for the simulation of biological organs. However, while codes and software packages for nonlinear FEM are readily available, codes and software packages for nonlinear BEM are not readily available. This author is not aware of any nonlinear BEM code that may be useful for the simulation of biological organs (e.g., 3D hyperelasticity). Even the commercial boundary element simulation software BEASY (developed by Prof. Brebbia who is widely considered to be the one who invented the BEM) does not include hyperelasticity. Developing one's own nonlinear BEM codes would require significant amount of time, resources, and expertise. These may be the reasons why no one has used nonlinear BEM for the real-time simulation of biological organs.

## 2. Literature on Nonlinear BEM

As regards to the use of the BEM for solving nonlinear problems (2D and 3D), one cannot find as much literature as one would expect to see. In fact, one can find only a few references on this topic. An attempt has been made in the following paragraphs to summarize important references on nonlinear BEM, especially those that are important from the point of view of the simulation of biological organs.

The reference [Wei-Zhe Feng, et al., 2015] presents a new BEM for solving 2D and 3D elastoplastic problems without initial stresses/strains. The reference [Trevor G. Davies and Xiao-Wei Gao, 2006] uses the boundary element method to carry out three-dimensional elasto-plastic analysis. The reference [Katia Bertoldi, et al., 2005] presents a new boundary element technique for elastoplastic solids. The technique does not use domain integrals. The reference [Xiao-Wei Gao and Trevor G. Davies, 2000] presents an effective boundary element algorithm for 2D and 3D elastoplastic problems.

The reference [P.M. Baiz and M.H. Aliabadi, 2007] analyzes the buckling of shear deformable shallow shells using the boundary element method, while [M.H. Aliabadi, 2006] uses the boundary element method to analyze shear deformable plates with combined geometric and material nonlinearities. The reference [T. Dirgantara and M.H. Aliabadi, 2006] uses a boundary element formulation to perform geometrically nonlinear analysis of shear deformable shells. The reference [P.H. Wen, et al., 2005] carries out large deflection analysis of Reissner plate using the boundary element method. The reference [Hui-Shen Shen, 2000] discusses the nonlinear bending of simply supported rectangular Reissner–Mindlin plates resting on elastic foundations under transverse and in-plane loads.

The reference [M. Brun, et al., 2003] discusses a boundary element technique for incremental, nonlinear elasticity.

Many a time biological organ may be assumed to be hyperelastic. As far as accuracy of the simulations is concerned (if speed is of no concern), one is expected to get more realistic results by assuming that biological organs are hyperelastic instead of assuming that they follow the linear elastostatic behaviour. Hence, sources from the literature that use the BEM for modelling hyperelasticity are identified in the next paragraph.

The reference [O. Köhler and G. Kuhn, 2001] discusses the application of the Domain-Boundary Element Method (DBEM) for solving hyperelastic and elastoplastic finite deformation problems (axisymmetric and 2D/3D problems). The reference [G. Karami and D. Derakhshan, 2001] introduces a field boundary element method for large deformation analysis of hyperelastic problems.

## 3. Nonlinear Formulations and Coding

The references [O. Köhler and G. Kuhn, 2001; G. Karami and D. Derakhshan, 2001] quite extensively and clearly describe the BEM formulation for hyperelasticity. These formulations may readily be used for the nonlinear BEM-based simulation of biological organs.

One of the differences between a nonlinear BEM code and a linear BEM code is that while carrying out nonlinear simulations, the characteristic matrix has to be calculated using a nonlinear formulation, e.g., the formulation explained in [O. Köhler and G. Kuhn, 2001; G. Karami and D. Derakhshan, 2001]. The other difference is that the system of algebraic equations to be solved to get the final solution is not linear. Hence a solution method that can solve a system of nonlinear algebraic equations has to be employed at the last stage.

Codes may be parallelized to make them run faster. Hardware that may be utilized include computer cluster, supercomputer, GPU, using Intel Many Integrated Core

Architecture (Intel MIC, which is a coprocessor), using a processor with many cores (e.g., Knights Landing, which is a standalone processor). From the reference [Victor W. Lee, et. al, 2010], it seems that GPUs may or may not be as good as they appear to be. However, a single processor with many cores is likely to be helpful for real-time simulations since the time for data transfer between computing cores for this type of processors is very small because all the cores are present in the same chip (same piece of semiconductor). One may also note that a few researchers are trying to develop processors that would have a few hundred cores each.

## 4. Comparison between the Results from Nonlinear Analysis (using ANSYS) and Linear Analysis (using ANSYS)

In this section, comparison is made between the results obtained by using linear elastostatic analysis and nonlinear analysis. The commercial software package ANSYS is used for the purpose; ANSYS is used for both linear and nonlinear analysis here.

The biological organ simulated in this section and the next section is the left kidney of the Visible Human male [VHP, n.d.]. The geometry of the kidney was obtained by following the procedure mentioned in [Kirana Kumara P, 2012]. The kidney is finally represented by *96* surface triangles.

Human kidneys are subjected to boundary conditions that are so complicated that it is virtually impossible to reproduce the boundary conditions in a computer model. Hence the kidney is subjected to arbitrary boundary conditions here. The idea is that if a computer model can give accurate solutions for many sets of arbitrary boundary conditions, then it is reasonable to assume that the user can specify whatever set of boundary conditions one wants to impose and the solution obtained for the specified set of boundary conditions would be accurate. The kidney considered in this work is subjected to three different sets of boundary conditions. Hence there are three problems to be solved.

In this work, Problem 1, Problem 2, and Problem 3 refer to cases where the kidney under consideration is simulated. The location where the kidney is fixed (i.e., zero displacement specified in all the x, y, and z directions), and the location where a specified non-zero displacement is specified are shown in Figure 1. For each of these three problems, element numbers *8*, *15*, and *24* are subjected to the zero-displacement condition in each of the x, y, and z directions, and the element number *94* is subjected to a non-zero displacement.

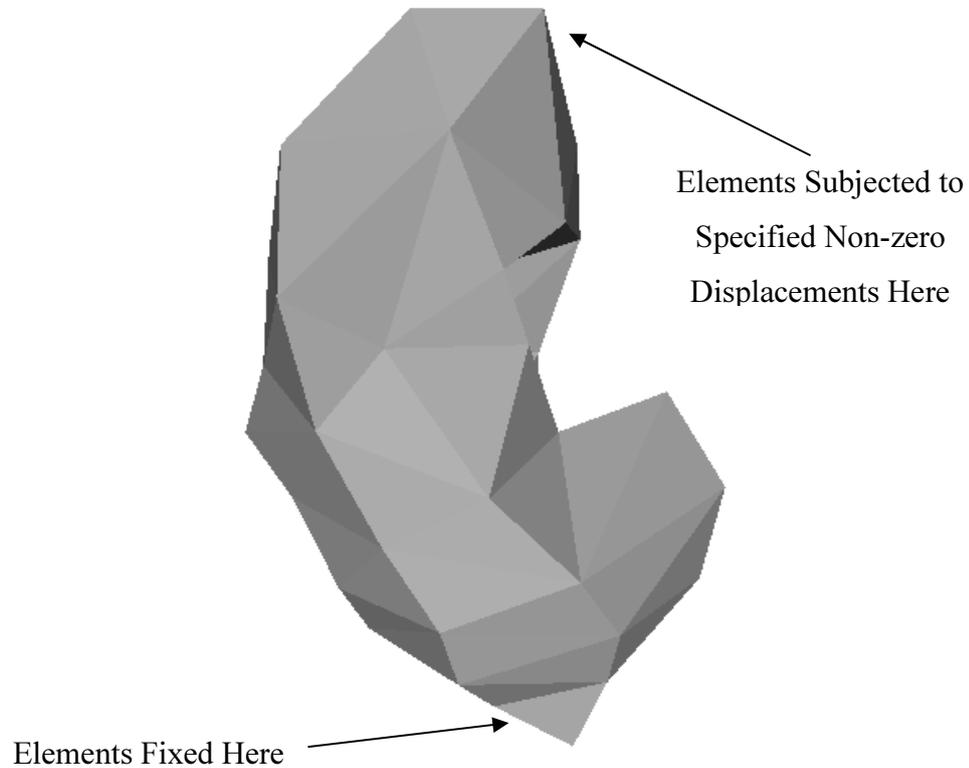

Elements Subjected to Specified Non-zero Displacements Here

Elements Fixed Here

**Figure 1** Boundary Conditions for the Kidney

For Problem 1, the element *94* is subjected to the non-zero displacement of *5 mm* in the x direction. For Problem 2, the element *94* is subjected to the non-zero displacement of *5 mm* in the y direction. For Problem 3, the element *94* is subjected to the non-zero displacement of *5 mm* in the z direction. The value of *5 mm* is chosen for all the problems because this value of displacement corresponds to about *5%* deformation along the largest dimension of the biological organ considered, if the load that causes the deformation is applied along the same direction. Of course, specifying *5 mm* displacement in other directions can result in deformations that are not close to the *5%* deformation. However, one can note that the idea here is to specify physically meaningful non-zero displacement boundary conditions. This author has not aligned the kidney to match the largest dimension of the kidney to any of the x, y, and z axes. Hence, although the non-zero displacement boundary conditions are specified along only one of x, y, and z axes (at a time) for all of the problems considered, one can note that it is reasonable to assume that the biological organ has been subjected to arbitray boundary conditions.

The material properties used are: Young's modulus = 150 N/mm$^2$, Poisson's ratio = 0.4, and only geometric nonlinearity is considered (the material is considered to be linear elastic, but the material can undergo large deformation). Literature says that it is much more important to incorporate geometric nonlinearity when compared to incorporating nonlinear material models, and many a times just incorporating geometric nonlinearity while using just the linear elastic constitutive model results in highly accurate simulations [Adam Wittek, et al., 2009].

Now, a comparison between linear and nonlinear analysis is done for the three simulations involving the kidney. The element type used is Tet 10node 187. The geometry is discretized into 782 nodes in total. The discretized geometry (undeformed geometry), as displayed in ANSYS, is shown in Figure 2. Figure 3 shows the undeformed and deformed geometry, displayed over the same frame, for Problem 1 and for the linear analysis (rendered mesh refers to the undeformed geometry whereas the wireframe mesh refers to the deformed geometry). Similarly, Figure 4 shows the undeformed and deformed geometry, displayed over the same frame, for Problem 1 and for the nonlinear analysis (rendered mesh refers to the undeformed geometry whereas the wireframe mesh refers to the deformed geometry).

Similarly, Figure 5 shows the undeformed and deformed geometry, for Problem 2 and for the linear analysis (For Figure 5 to Figure 8, rendered mesh refers to the undeformed geometry whereas the wireframe mesh refers to the deformed geometry). Figure 6 shows the undeformed and deformed geometry, for Problem 2 and for the nonlinear analysis. Figure 7 shows the undeformed and deformed geometry, for Problem 3 and for the linear analysis. Figure 8 shows the undeformed and deformed geometry, for Problem 3 and for the nonlinear analysis.

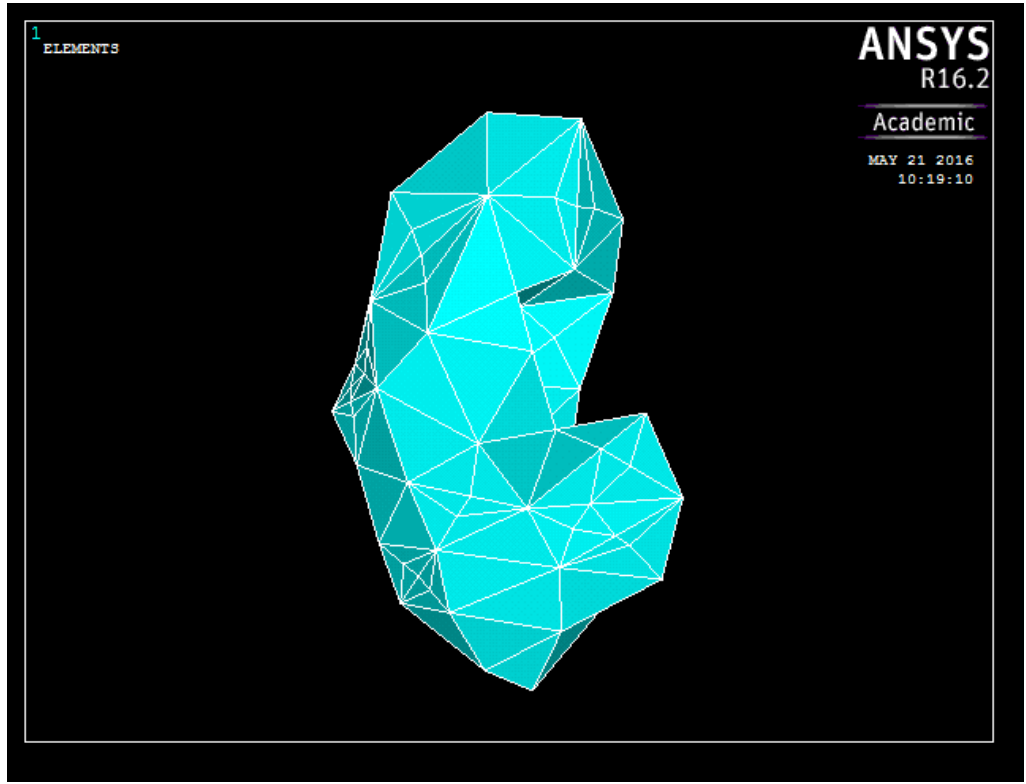

**Figure 2** Discretized Geometry as Displayed in ANSYS

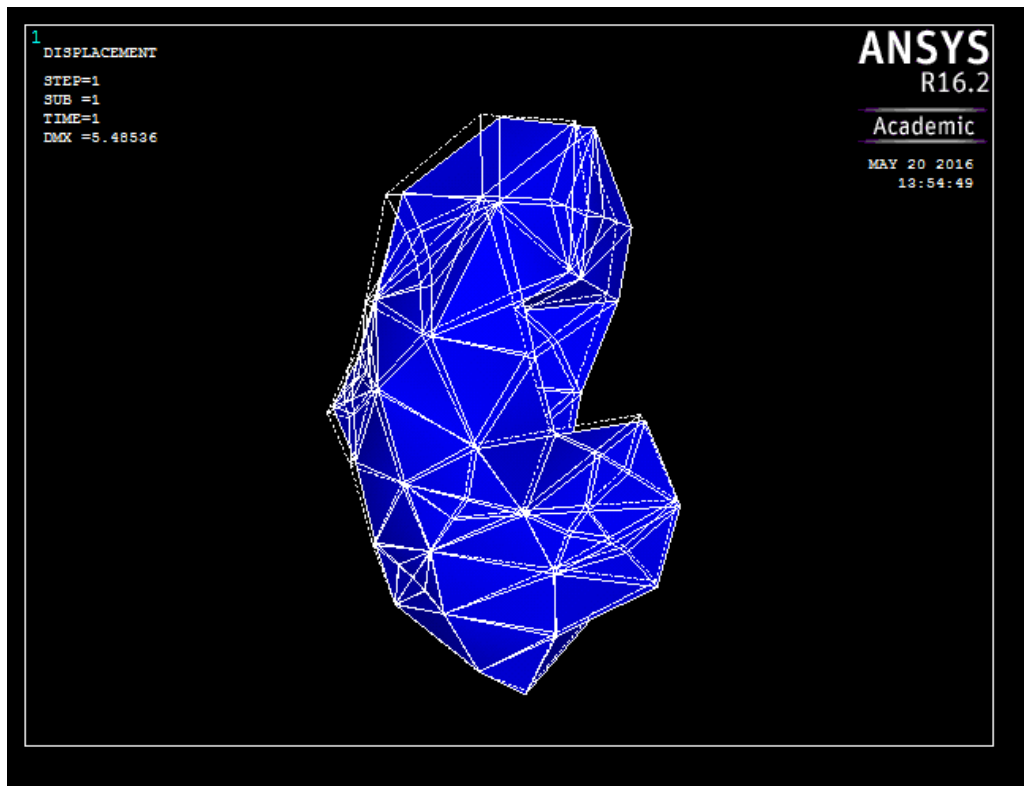

**Figure 3** Undeformed and Deformed Geometry for Problem 1 (Linear Analysis)

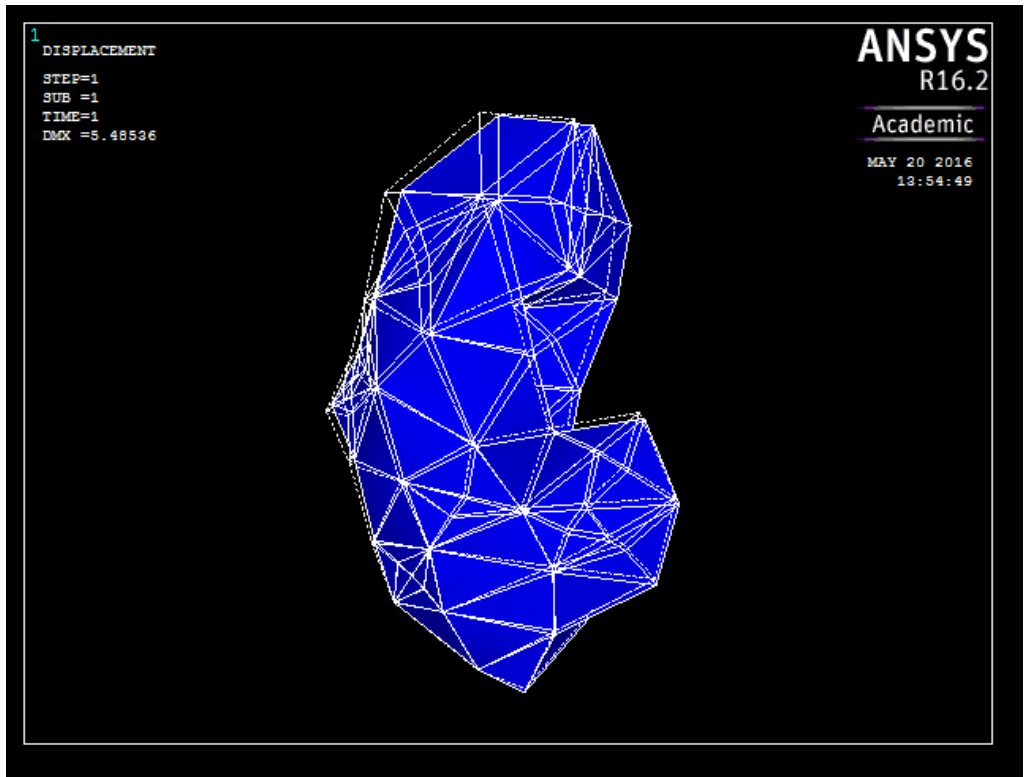

**Figure 4** Undeformed and Deformed Geometry for Problem 1 (Nonlinear Analysis)

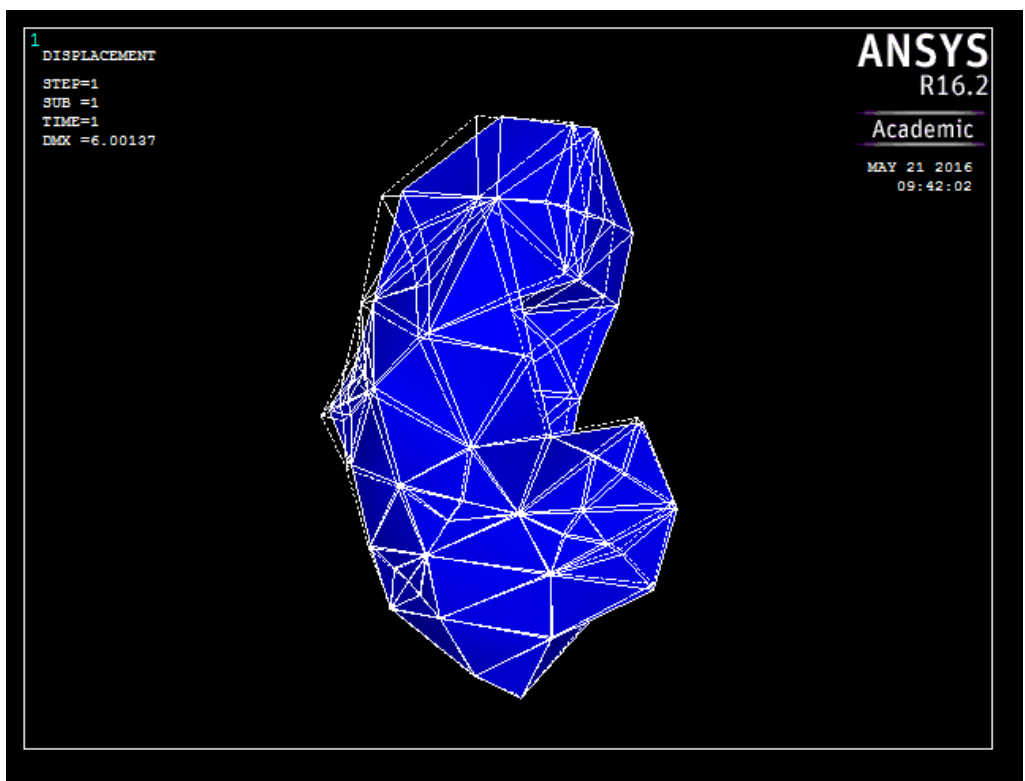

**Figure 5** Undeformed and Deformed Geometry for Problem 2 (Linear Analysis)

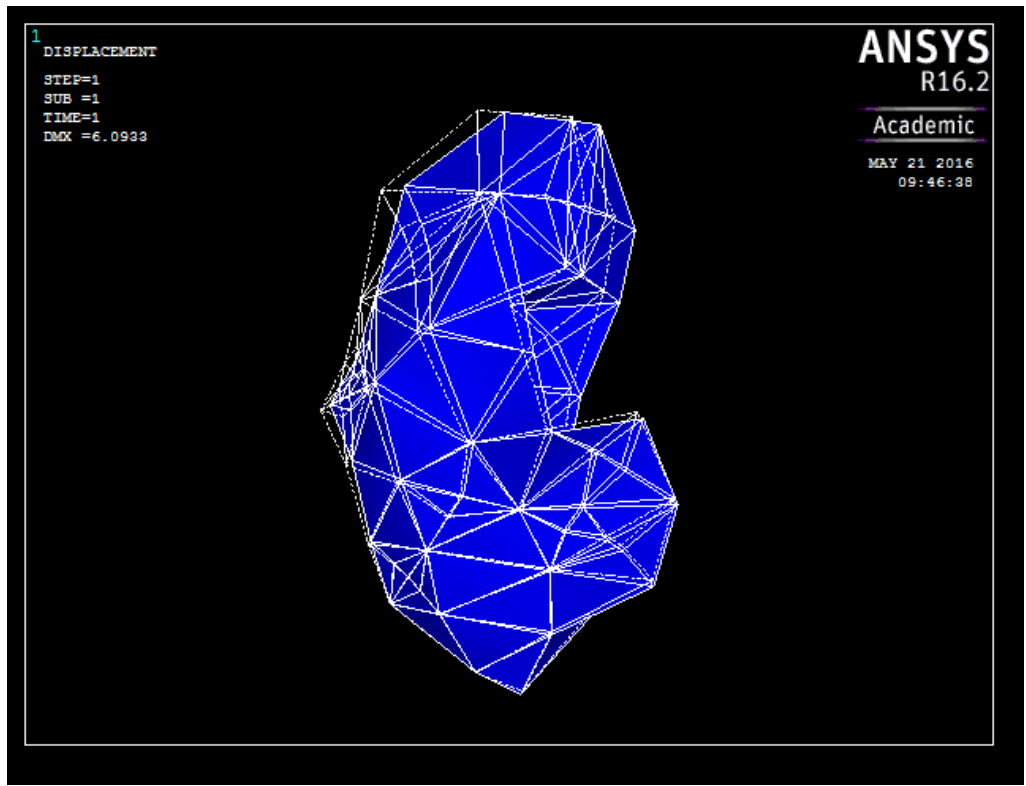

**Figure 6** Undeformed and Deformed Geometry for Problem 2 (Nonlinear Analysis)

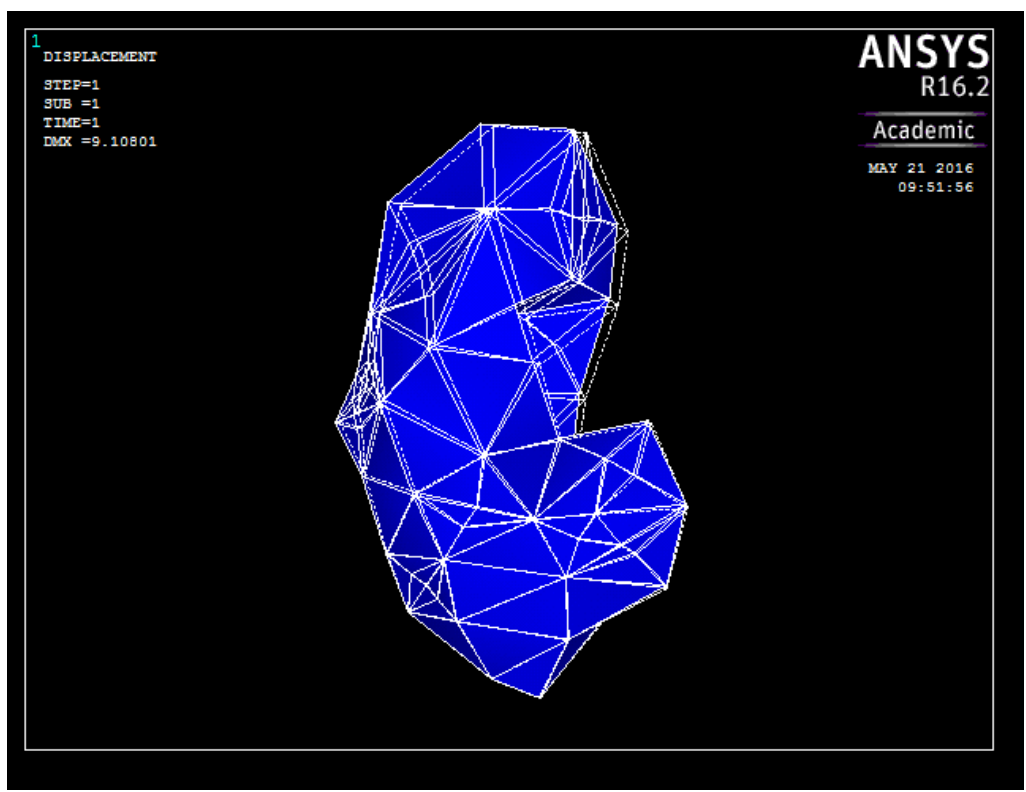

**Figure 7** Undeformed and Deformed Geometry for Problem 3 (Linear Analysis)

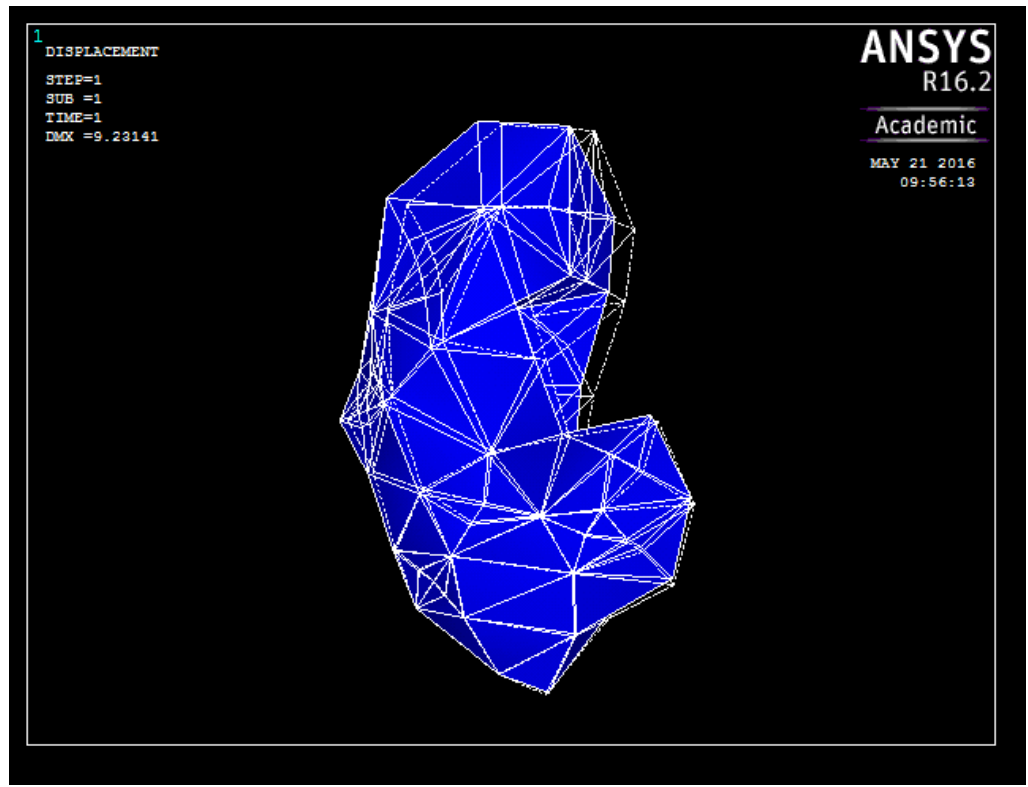

**Figure 8** Undeformed and Deformed Geometry for Problem 3 (Nonlinear Analysis)

It is easier to compare the difference between the results from linear and nonlinear analyses if the actual values of the displacements are tabulated. Hence, for each of the analyses above, the values of the displacement vector sum at eleven distinct nodes is noted down. These values are tabulated in Table 1 to Table 3. The eleven nodes are selected such that they are not from a certain part of the geometry only (i.e., nodes are scattered throughout the geometry). The node numbers of the chosen nodes are: 41, 43, 50, 49, 34, 15, 11, 4, 18, 246, 20.

**Table 1** Percentage Errors for Problem 1

| Node Number | Displacement Vector Sum for the Large Deformation Analysis *(mm)* | Displacement Vector Sum for the Small Deformation Analysis *(mm)* | Percentage Error |
|---|---|---|---|
| 41 | 4.176 | 5.011 | 19.995 |
| 43 | 5.485 | 5.983 | 9.079 |
| 50 | 5.161 | 5.854 | 13.428 |
| 49 | 5.177 | 5.210 | 0.637 |
| 34 | 2.200 | 2.300 | 4.545 |
| 15 | 0.000 | 0.000 | 0.000 |
| 11 | 0.728 | 0.659 | -9.478 |
| 4 | 0.750 | 0.982 | 30.933 |
| 18 | 1.384 | 1.375 | -0.650 |
| 246 | 1.551 | 1.583 | 2.063 |
| 20 | 1.870 | 1.763 | -5.722 |

**Table 2** Percentage Errors for Problem 2

| Node Number | Displacement Vector Sum for the Large Deformation Analysis *(mm)* | Displacement Vector Sum for the Small Deformation Analysis *(mm)* | Percentage Error |
|---|---|---|---|
| 41 | 5.043 | 5.011 | -0.635 |
| 43 | 6.048 | 5.983 | -1.075 |
| 50 | 5.932 | 5.854 | -1.315 |
| 49 | 5.251 | 5.210 | -0.781 |
| 34 | 2.323 | 2.300 | -0.990 |
| 15 | 0.000 | 0.000 | 0.000 |
| 11 | 0.661 | 0.659 | -0.303 |
| 4 | 0.984 | 0.982 | -0.203 |
| 18 | 1.373 | 1.375 | 0.146 |
| 246 | 1.584 | 1.583 | -0.063 |
| 20 | 1.767 | 1.763 | -0.226 |

Table 3 Percentage Errors for Problem 3

| Node Number | Displacement Vector Sum for the Large Deformation Analysis *(mm)* | Displacement Vector Sum for the Small Deformation Analysis *(mm)* | Percentage Error |
|---|---|---|---|
| 41 | 7.551 | 7.260 | -3.854 |
| 43 | 8.922 | 8.726 | -2.197 |
| 50 | 9.231 | 9.108 | -1.332 |
| 49 | 8.446 | 8.467 | 0.249 |
| 34 | 3.146 | 3.090 | -1.780 |
| 15 | 0.000 | 0.000 | 0.000 |
| 11 | 0.948 | 0.910 | -4.008 |
| 4 | 1.488 | 1.371 | -7.863 |
| 18 | 1.610 | 1.459 | -9.379 |
| 246 | 1.776 | 1.556 | -12.387 |
| 20 | 2.051 | 1.822 | -11.165 |

## 5. Comparison between the Results from Nonlinear Analysis (using ANSYS) and Linear Analysis (using this Author's BEM Code)

Instead of comparing the results from nonlinear analysis using ANSYS with the results from linear analysis using ANSYS (as done in the last section), it is better if the linear analysis is conducted using this author's BEM code [Kirana Kumara P, 2014] (instead of ANSYS). However, here one needs to compare the results at the same location (node) of the problem domain, and it is difficult to do this since FEM and BEM use different discretizations and hence it is difficult to have nodes at the same locations for both the discretizations. Still a comparison is done in this section by manually (visually) locating the nodes in the BEM discretization, which may be located approximately at the same location as the corresponding nodes in the finite element model (ANSYS). Of course, it may be noted that the nodes in the BEM discretization are not located exactly at the same location as the corresponding nodes in the FEM discretization, and this itself can be a cause of some amount of error.

The same Problem 1, Problem 2, and Problem 3 chosen for the simulations for the last section are chosen for the simulations for this section also. Of course, for each of Problem 1, Problem 2, and Problem 3 here, geometry, loads and boundary conditions, and material properties (both for the linear analysis and the nonlinear analysis) used here are the same as the ones mentioned in the last section.

The results from linear and nonlinear analyses are compared by tabulating the actual values of the displacement vector sum at eleven distinct points (tables like those in the last section). The values are tabulated in Table 4 to Table 6. The eleven nodes in the FEM model are the same as those chosen in the last section, i.e., 41, 43, 50, 49, 34, 15, 11, 4, 18, 246, 20. The eleven nodes in the BEM model are chosen such that they are located approximately at the same location as the corresponding FEM nodes, and the node numbers of the corresponding BEM nodes are: 84, 87, 96, 83, 68, 24, 13, 11, 22, 25, 50.

**Table 4** Percentage Errors for Problem 1

| Node Number | Displacement Vector Sum for the Large Deformation Analysis *(mm)* | Displacement Vector Sum for the Small Deformation Analysis *(mm)* | Percentage Error |
|---|---|---|---|
| 84 | 4.176 | 3.911 | -6.346 |
| 87 | 5.485 | 4.513 | -17.721 |
| 96 | 5.161 | 4.932 | -4.437 |
| 83 | 5.177 | 4.133 | -20.166 |
| 68 | 2.200 | 1.904 | -13.455 |
| 24 | 0.000 | 0.000 | 0.000 |
| 13 | 0.728 | 0.852 | 17.033 |
| 11 | 0.750 | 0.933 | 24.400 |
| 22 | 1.384 | 1.549 | 11.922 |
| 25 | 1.551 | 1.625 | 4.771 |
| 50 | 1.870 | 1.723 | -7.861 |

**Table 5** Percentage Errors for Problem 2

| Node Number | Displacement Vector Sum for the Large Deformation Analysis *(mm)* | Displacement Vector Sum for the Small Deformation Analysis *(mm)* | Percentage Error |
|---|---|---|---|
| 84 | 5.043 | 3.025 | -40.016 |
| 87 | 6.048 | 3.826 | -36.739 |
| 96 | 5.932 | 4.322 | -27.141 |
| 83 | 5.251 | 4.115 | -21.634 |
| 68 | 2.323 | 2.404 | 3.487 |
| 24 | 0.000 | 0.000 | 0.000 |
| 13 | 0.661 | 0.150 | -77.307 |
| 11 | 0.984 | 0.162 | -83.537 |
| 22 | 1.373 | 0.389 | -71.668 |
| 25 | 1.584 | 0.546 | -65.530 |
| 50 | 1.767 | 0.839 | -52.518 |

Table 6 Percentage Errors for Problem 3

| Node Number | Displacement Vector Sum for the Large Deformation Analysis *(mm)* | Displacement Vector Sum for the Small Deformation Analysis *(mm)* | Percentage Error |
|---|---|---|---|
| 84 | 7.551 | 3.707 | -50.907 |
| 87 | 8.922 | 3.775 | -57.689 |
| 96 | 9.231 | 3.838 | -58.423 |
| 83 | 8.446 | 3.389 | -59.874 |
| 68 | 3.146 | 1.717 | -45.423 |
| 24 | 0.000 | 0.000 | 0.000 |
| 13 | 0.948 | 0.747 | -21.203 |
| 11 | 1.488 | 0.792 | -46.774 |
| 22 | 1.610 | 0.999 | -37.950 |
| 25 | 1.776 | 1.096 | -38.288 |
| 50 | 2.051 | 1.183 | -42.321 |

**6. Discussion and Concluding Remarks**

From the tables and the figures in the last two sections, one may see that there is not too much difference between the results obtained by linear analysis and nonlinear analysis for many of the cases, especially when ANSYS is used both for linear analysis and nonlinear analysis although the percentage errors might be significant when this author's code is used for the linear analysis. However, it is difficult to definitively say how much error is allowed. Only surgeons can say whether a simulation is useful or not. In fact, validating a numerical model by taking feedback from many surgeons, many surgical procedures, and many trials could itself be a research topic. As of present, there is no clarity on what the allowable error in a simulation is, and the subject is a research topic which has not been explored well.

This author's stand is that it is good to stick to linear elastostatic behaviour at present, and as more powerful hardware (together with relevant software) becomes readily available in the future, it may be good to incorporate nonlinearity. Of course, the relevant technology (e.g., developing nonlinear BEM codes) may be developed right now, and possibly the results could be used with benefit in cases where there is no need for the simulations to be strictly real-time (e.g., surgery planning).

Xiao-Wei Gao, Trevor G. Davies, 2000, An effective boundary element algorithm for 2D and 3D elastoplastic problems, International Journal of Solids and Structures 37(36), p. 4987-5008